\documentclass[12pt]{article}
\usepackage[utf8]{inputenc}

\usepackage[margin=1in]{geometry} 
\usepackage{amsmath,amsthm,amssymb,amsfonts}
\usepackage{graphicx}
\usepackage{setspace}
\usepackage{sectsty}
\usepackage{thmtools, thm-restate, enumitem}
\usepackage{natbib}
\usepackage{color}
\usepackage{multirow}

\usepackage{soul}

\usepackage[flushleft]{threeparttable}
\usepackage{booktabs}

\usepackage{lipsum}
\usepackage{ragged2e}
\usepackage[affil-it]{authblk}

\allowdisplaybreaks

\title{Inference from Non-Random Samples Using Bayesian Machine Learning}

\author{Yutao Liu, Andrew Gelman, Qixuan Chen\thanks{Yutao Liu is a Ph.D. candidate, Andrew Gelman is a Professor of Statistics and Political Science, and Qixuan Chen is an Associate Professor of Biostatistics, at Columbia University.  Address correspondence to: Qixuan Chen, Department of Biostatistics, Columbia University, 722 West 168th Street, New York, NY 10032; E-mail: qc2138@cumc.columbia.edu. }  }

\date{}

\begin{document}
\maketitle

\vspace{-.5in}
\begin{abstract}
 We consider inference from non-random samples in data-rich settings where high-dimensional auxiliary information is available both in the sample and the target population, with survey inference being a special case. We propose a regularized prediction approach that predicts the outcomes in the population using a large number of auxiliary variables such that the ignorability assumption is reasonable while the Bayesian framework is straightforward for quantification of uncertainty. Besides the auxiliary variables, inspired by \citet*{little2004robust}, we also extend the approach by estimating the propensity score for a unit to be included in the sample and also including it as a predictor in the machine learning models. We show through simulation studies that the regularized predictions using soft Bayesian additive regression trees yield valid inference for the population means and coverage rates close to the nominal levels. We demonstrate the application of the proposed methods using two different real data applications, one in a survey and one in an epidemiology study.  \\
 
  
\end{abstract}

\maketitle


%

\raggedright
\setlength{\parskip}{\baselineskip}

\section{Introduction}
\label{s:intro}

Inference about a target population based on sample data relies on the assumption that the sample is representative. However, simple random samples are often not available in real data problems. Therefore, there is a need to generalize inference from the available non-random sample to the target population of interest. For example, randomized controlled trials (RCTs) are considered a gold standard to estimate treatment effects, but the measured effects can only be formally generalized to the participants within the trial. Recent evidence has indicated that subjects in an RCT can be much different from patients in routine practice. Such concern among clinicians about the external validity of RCTs has led to the underuse of effective treatments \citep{rothwell2005external}. This highlights the importance of generalizing treatment effect of RCTs to a definable patient population.

Survey sampling is a field that specifically deals with inference on populations with non-random samples, which can be viewed as a special case of generalizing inference. Probability samples collected via probability surveys have historically proven effective. However, such data comes with considerable cost, both time and budget. In the past several decades, large scale probability surveys have suffered increasingly high non-response rates, besides the rising costs. The probability surveys with low response rates are often non-representative, which challenges the validity of survey inference.  In the meanwhile, recent development of information technology makes it increasingly convenient and cost-effective to collect large numbers of samples with detailed information via online surveys and opt-in panels. Such samples are highly non-representative due to selection bias. Classical weighting methods in survey literature such as post-stratification \citep{valliant1993poststratification} and raking \citep{deming1940least} can improve representativeness of survey samples when a small number of discrete auxiliary variables about populations are available for survey adjustments. However, such weighting methods can yield highly variable estimates of population quantities in the presence of extreme weights. Alternatively, model-based methods can be used. \citet*{wang2015forecasting} demonstrates, through election forecast with non-representative voter intention polls on the Xbox gaming platform, that multilevel regression and post-stratification (MRP)
 can be used to generate accurate survey estimates from non-representative samples.   Their estimates are in line with the forecasts from leading poll analyst. MRP is very appealing when statistical adjustment are made using a small number of discrete auxiliary variables.

In recent years, population data of high volume, variety, and velocity has become increasing available, with examples including administrative data or electronic medical records. Such data contained detailed individual level information with high-dimensionality and can be used to generalize inference of non-random samples to their target populations. Although post-stratification, raking, and MRP methods can improve representativeness in the presence of a small number of discrete auxiliary variables, they are infeasible to be applied in high-dimensional settings. 
With high-dimensional auxiliary variables, Bayesian machine learning techniques have been shown to be effective in improving statistical inference in missing data and causal inference \citep*{hill2011bayesian,tan2019robust,hahn2020bayesian}. Specially, \cite{hill2011bayesian} shows that Bayesian additive regression trees (BART) produces more accurate estimates of average treatment effects compared to propensity score matching, propensity-weighted estimators, and regression adjustment when the response surface is nonlinear and not parallel between treatment and control groups. \citet*{tan2019robust} demonstrate, in the presence of missing data, that BART reduces bias and root mean square error of the doubly robust estimators when both propensity and mean models were misspecified. Inspired by these works, we propose Bayesian machine learning model-based methods and extensions for estimating population means using non-random samples. The proposed methods can be applied not only in the context of survey inference but also in more general settings, such as RCTs and epidemiological observational studies. We evaluate the proposed methods using  simulation studies and demonstrate their applications in a mental health survey of Ohio Army National Guard service members and a non-random sample from an observational study using electronic medical records of COVID-19 patients.

\section{Methods}

\subsection{Notation and background}

\begin{figure}
    \centering
    \includegraphics[width=\linewidth]{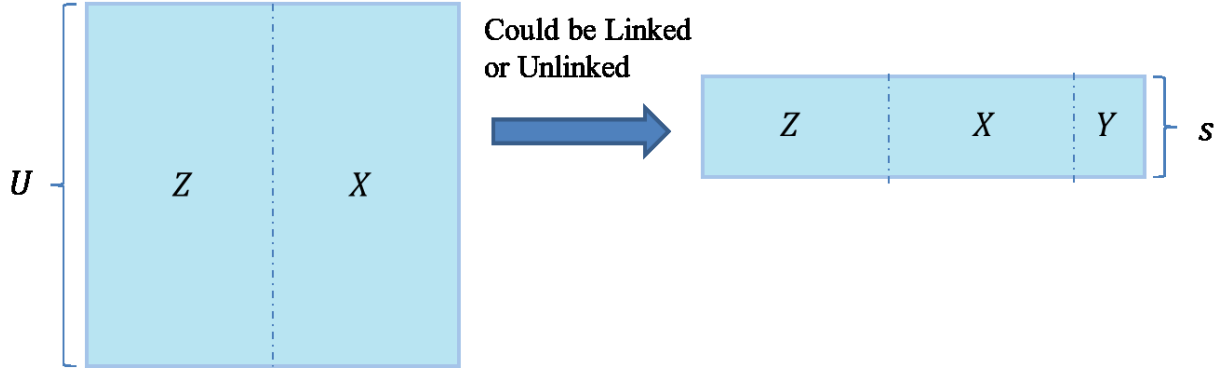}
    \caption{Population $U$ and non-random sample $s$ with shared discrete auxiliary variables $\boldsymbol{Z}$ and continuous auxiliary variables $\boldsymbol{X}$ as well as outcome $Y$ measured only in $s$.}
    \label{notation}
\end{figure}

Let $U$ be the finite population of size $N$ and $s$ be a non-random sample of size $n$ from the population. In the sample $s$, information on the outcome of interest $Y$, discrete auxiliary variables $\boldsymbol{Z}$ and continuous auxiliary variables $\boldsymbol{X}$ were collected. In addition, data from the population $U$ (e.g. census, administrative data, or electronic medical records) is also available with the same set of auxiliary variables $\boldsymbol{Z}$ and $\boldsymbol{X}$ measured for all units in the population. Figure~\ref{notation} illustrates the scenario under consideration, with population data on the left and the sample data on the right. Without loss of generality, we consider a continuous variable of interest $Y$ with the estimand of interest being the finite population mean $Q(Y) = \frac{1}{N}\sum_{i \in U} Y_i$. 

When the dimensions of $\boldsymbol{Z}$ and $\boldsymbol{X}$ are small, post-stratification, raking, and MRP can be applied by first discretizing the continuous auxiliary variables $\boldsymbol{X}$ as $\boldsymbol{X}^*$ using quantiles. Using the joint distribution of discrete auxiliary variables $(\boldsymbol{Z}, \boldsymbol{X}^{*})$, \emph{post-stratification} partitions the population into $J$ disjoint post-strata with $U = \bigcup_{j = 1}^{J} U_j$ of size $N_j$ and the sample into subsamples with $s = \bigcup_{j = 1}^{J} s_j$ of size $n_j$ for the $j$th post-stratum, correspondingly. With respect to the post-strata, the finite population mean can be rewritten as $$Q(Y) = \frac{1}{N} \sum_{j = 1}^J \sum_{i \in U_j} Y_{i} = \frac{1}{N} \sum_{j = 1}^J N_j \theta_j,$$ where $\theta_j = \frac{1}{N_j} \sum_{i \in U_j} Y_i$ is subpopulation mean of post-stratum $U_j$. With the assumption that the  
sample units in each post-stratum are representative of population units in that post-stratum, the post-strata means are estimated using corresponding subsample means $\hat{\theta}_j = \bar{y}_j = \frac{1}{n_j} \sum_{i \in s_j} y_i$. Naturally, the post-stratification (PS) estimator takes the form \begin{equation}
    \widehat{Q}_{\text{PS} } = \frac{1}{N} \sum_{j = 1}^J N_j \bar{y}_j = \frac{1}{N} \sum_{i \in s} w_{i} y_i, \label{PS}
\end{equation} where $w_{i} = N_j / n_j$ for $i \in U_j$ is the post-stratification weight assigned to sample unit $i$ in post-stratum $j$ which is inverse proportional to the sampling fraction $n_j / N_j$. The post-stratification estimator could be numerically unstable when such partition results in small cells in the sample, in other words, small $n_j$ and large weights $w_j$.

Alternatively, \emph{raking} generates weights $w_i$ to match successively the marginal (rather than the joint) distributions of $(\boldsymbol{Z}, \boldsymbol{X}^{*})$ via iterative proportional fitting. 
The raking weighted estimator takes the form \begin{equation}
    \widehat{Q}_{  \text{R} } = \frac{1}{N} \sum_{i \in s} w_i {y_i}, \label{rake}
\end{equation}  with $w_i$ denoting raking weights. Raking weights could be highly variable, so the resulting weighted estimators could be inefficient. Also, raking may have convergence issues as the number of auxiliary variables increases. 

\citet*{gelman2007struggles} reviews a model-based perspective on the PS estimator. In the model-based approach, a regression model is specified to model the conditional distribution of outcome given the discrete auxiliary variables $p(Y | \boldsymbol{Z}, \boldsymbol{X}^{*})$.  Define stratum-specific means $\theta_j = \text{E}(Y_i | \boldsymbol{Z}_i, \boldsymbol{X}^{*}_i)$, $i \in U_j$. And estimating $\hat{\theta}_j = \widehat{\text{E}}(Y_i | \boldsymbol{Z}_i, \boldsymbol{X}^{*}_i)$ based on the fitted model leads to the  \emph{regression and post-stratification} (RP) estimator  \begin{equation}
    \widehat{Q}_{\text{RP} } = \frac{1}{N} \sum_{j = 1}^J N_j \hat{\theta}_j. \label{RP}
\end{equation} As a special case, specifying a saturated regression model (including all possible interactions terms) allows $J$ post-stratum specific means and the least square estimators  $\hat{\theta}_j = \bar{y}_j = \frac{1}{n_j} \sum_{i \in s_j} y_i$. As a result, $\widehat{Q}_{\textrm{RP} } = \widehat{Q}_{\textrm{PS} }$.

From the model-based perspective, the problem of unstable estimates due to small cells in post-stratification can be viewed as a model fitting problem due to model complexity. Such perspective motivates using alternative modeling techniques to improve estimation. Instead of using classical saturated regression models, \emph{multilevel regression and post-stratification} (MRP) utilizes hierarchical regression models to achieve stable estimates.  Both main effects and interaction terms could be specified as multilevel random effects so that information across post-strata can be partially pooled in the model fitting procedure \citep{gelman1997poststratification}. MRP improves efficiency in the population mean estimation than post-stratification and raking when data are sparse in some post-strata. 

Still, it is challenging to perform MRP in high-dimensional setting, especially in the presence of a large number of noise variables not associated with $Y$, because a parametric form needs to be specified for the multilevel regression. Also, continuous auxiliary variables need to be discretized before modeling.

The model-based RP approach can also be viewed as a prediction approach and the RP estimator in (\ref{RP}) can be rewritten as $$\widehat{Q}_{\textrm{RP} } = \frac{1}{N} \sum_{j = 1}^J N_j \hat{\theta}_j = \frac{1}{N} \sum_{j = 1}^J \sum_{i \in U_j} \hat{\theta}_j  = \frac{1}{N} \sum_{j = 1}^J \sum_{i \in U_j} \widehat{\textrm{E}}(Y_i | \boldsymbol{Z}_i, \boldsymbol{X}^{*}_i)  = \frac{1}{N} \sum_{i \in U} \widehat{\textrm{E}}(Y_i | \boldsymbol{Z}_i, \boldsymbol{X}^{*}_i), $$ where $\widehat{\textrm{E}}(Y_i | \boldsymbol{Z}_i, \boldsymbol{X}^{*}_i)$ is predictive value of $Y_i$ based on model $p(Y | \boldsymbol{Z}, \boldsymbol{X}^{*})$. Such perspective motivates the use of modern statistical techniques for generalization of inference via valid predictions of the outcomes in the population.  Specifically, the classical regression models in \emph{regression and post-stratification} can be replaced by any regularized prediction methods that achieve stable estimates while including high-dimensional covariates. Such models also allows modeling the continuous $\boldsymbol{X}$ directly. 

\subsection{New Approach: Regularized Prediction}

Tree-based methods are appealing techniques for handling high-dimensional problems. Sum-of-trees ensembles achieve high prediction accuracy and better approximate the functional forms of continuous variables, with each single tree regularized to obtain stable predictions and achieve bias variance trade-off. Taking a model-based predictive perspective, we extend the RP approach to high-dimensional setting by replacing parametric regression models with regularized additive regression trees. We consider the Bayesian modeling framework, as it is natural to implement predictive inference and straightforward for quantification of uncertainty.

\subsubsection{ BART and Soft BART Prediction}\label{sec:bart}

In the current setting, the conditional distribution of a continuous outcome given the high-dimensional auxiliary variables  $p(Y | \boldsymbol{Z}, \boldsymbol{X})$ can be modeled using Bayesian additive regression trees (BART) \citep*{chipman2010bart} or soft Bayesian additive regression trees (SBART) \citep*{linero2018bayesian}.

For continuous outcomes, BART and SBART assume Gaussian noise and model the location parameter using a non-parametric sum-of-trees structure, allowing both discrete and continuous auxiliary variables \begin{align} \label{BART-P}
    Y = G(\boldsymbol{Z}, \boldsymbol{X}) + \epsilon = \sum_{m = 1}^M g(\boldsymbol{Z}, \boldsymbol{X}; T_m, \boldsymbol{\mu}_m) + \epsilon, \quad \epsilon \overset{ \textrm{i.i.d.} }{\sim} N(0, \sigma^2), 
\end{align} where $M$ is fixed number of trees in the sum-of-trees structure, $T_m$ is the $m$-th binary tree with $\boldsymbol{\mu}_m$ being the parameters associated with the terminal nodes, and $g(\cdot)$ is the function assigning $\boldsymbol{\mu}_m$ according to $(\boldsymbol{Z}, \boldsymbol{X})$.  The sum-of-trees structure naturally handles high-dimensional auxiliary variables without specifying a parametric form, accounting for categorical variables, continuous variables and possible interactions. In the Bayesian framework, quantification of uncertainty is naturally characterized by the posterior and posterior predictive distributions.

In BART, $g(\cdot)$ is a deterministic function and the potential effect of continuous predictors, either linear or nonlinear, is approximated by step functions generated by cutting the continuous predictor at various splitting points in different trees. Regularization priors are specified on $p(T_m)$, $p(\boldsymbol{\mu}_m | T_m)$, $p(\sigma^2)$ such that each single tree $T_m$ is a weak learner. Such specification aims at preventing the individual tree effects from unduly influential and achieving stable predictions, with automatic default specifications facilitating easy implementation. For $p(T_m)$, the prior is specified by three aspects: (i) the probability that a node is nonterminal, (ii) the distribution on the splitting variable assignments at each interior node, and (iii) the distribution on the splitting rule assignment in each interior node, conditional on the splitting variable. For $p(\boldsymbol{\mu}_m | T_m)$ and $p(\sigma^2)$, conjugate normal distributions and inverse chi-square distributions are specified. 

In SBART, $g(\cdot)$ associates the values of  covariates with a probabilistic (instead of deterministic as in BART) path down the tree, with certain probability going left at each node. With such modification, a particular set of values of $(\boldsymbol{Z}, \boldsymbol{X})$ is associated with a certain terminal node with certain probability, obtained by averaging over all possible paths. Unlike hard decision trees in BART where each terminal node is constrained to influence the regression function locally, the soft decision trees in soft BART allow each terminal node to impose a global effect on the function. This global effect of local terminal nodes enables the soft decision trees to borrow information adaptively across different covariate regions. Sparsity-inducing priors are specified to achieve a balance between sparse and non-sparse settings.




In practice, cross validation could be applied to determine the number of trees $M$ and the hyperparameters in the Bayesian priors. \citet{linero2018bayesian} develop default prior specification with $M = 50$ which performs universally well in all the 10 benchmark datasets considered in the paper. In this paper, we consider $M = 50$ for BART and SBART in the simulation studies and applied examples. The BART and SBART prediction estimators of finite population mean, $\widehat{Q}_{ \textrm{BART}}$ and $\widehat{Q}_{ \textrm{SBART}}$, are obtained with the following steps. \begin{description}
\item [Step 1] Model $p(Y | \boldsymbol{Z}, \boldsymbol{X})$ using BART or soft BART,   $Y = G(\boldsymbol{Z}, \boldsymbol{X}) + \epsilon , \epsilon \sim N(0, \sigma^2)$ with corresponding Bayesian priors.
\item [Step 2] Obtain posterior distributions of $Q(Y) = \frac{1}{N} \sum_{i \in U} y_i$ using Markov chain Monte Carlo (MCMC) simulations. Specifically, in MCMC iteration $t$,  

\begin{enumerate}
            \item draw $G^{(t)}, \sigma^{(t)}  | Y_{i \in s}, \boldsymbol{Z}_{i \in U}, \boldsymbol{X}_{i \in U}$
            \item compute $\tilde{\theta}_i^{(t)} = G^{(t)} (\boldsymbol{Z}_i, \boldsymbol{X}_i)$ for $i \in U$
            \item obtain $\widehat{Q}_{ \textrm{(S)BART}}^{(t)} = \frac{1}{N} \left[ \sum_{i \in U}  \tilde{\theta}_i^{(t)}   + \left (\sum_{i \in s} y_i - \sum_{i \in s} \tilde{\theta}_i^{(t)} \right ) \right]$, using the observed $y_i$ in the sample and the predicted values for the population units that are not in the sample.
        \end{enumerate}
\item [Step 3] Obtain $\widehat{Q}_{ \textrm{(S)BART}}$: point estimates using (posterior) median of $\widehat{Q}_{ \textrm{(S)BART}}^{(t)}$ with credible intervals constructed using quantiles splitting the tails of posterior distribution equally.
\end{description}

In some cases, inference on subpopulation means  are also of interest, which can be obtained via modification of item 3 in Step 2, restricting the average to predictions and observed outcomes in the corresponding subpopulation  $\Omega \subset U$ and subsamples $s \cap \Omega$, $\widehat{Q}_{\Omega, \textrm{(S)BART}}^{(t)} = \frac{1}{N_{\Omega}} \left[ \sum_{i \in \Omega}  \tilde{\theta}_i^{(t)}   + \left (\sum_{i \in s \cap \Omega } y_i - \sum_{i \in s \cap \Omega } \tilde{\theta}_i^{(t)} \right ) \right]$ .

\subsubsection{BART and Soft BART Propensity Prediction}

 In the missing data literature, \citet*{little2004robust} proposed including logit-transformed response propensity score as covariates using splines in the imputation models. This response propensity prediction method yields robust estimates of sample means when the imputation model is misspecified. \citet*{tan2019robust} extended the method of \citet*{little2004robust} by using BART to fit both the imputation model and the response propensity model. They show that adding BART-estimated propensity score in the BART imputation model reduces bias and RMSE and improves confidence interval coverage rates in the mean estimation.

 Inspired by this,  we extend the BART and SBART prediction with a two-step approach. First, we estimate sample inclusion propensity using a propensity model. If the sample data are linked to the population data, we code the sample inclusion indicators $I=1$ for the units in the sample and $I=0$ for the rest of the units in the population. The propensity score $\hat{\pi}$ can then be estimated via modeling $p(I | \boldsymbol{Z}, \boldsymbol{X})$ using probit Bayesian additive regression trees \citep{chipman2010bart}. If the sample data is unlinked to the population data, we round the continuous $\boldsymbol{X}$ to $\left[ \boldsymbol{X} \right]$ at a certain precision level and identify $K$ categories with unique values of $(\boldsymbol{Z},\left[ \boldsymbol{X} \right])$. Within each category $k = 1, \ldots, K$, the number of units in the population $N_k$ and that in the sample $n_k$ can be counted. 
Once the counts $(N_k, n_k)$ are created for each category, the propensity score $\hat{\pi}$ for the units to be included in the sample, given $(\boldsymbol{Z}, \left[ \boldsymbol{X} \right])$,  can be obtained via models for binomial outcomes. 
Next, we model $p(Y | \boldsymbol{Z}, \boldsymbol{X}, \hat{\pi})$ by additionally including $\hat{\pi}$ as a covariate in BART or SBART model with the rest of the steps being the same as Section~\ref{sec:bart}. The detailed steps of obtaining the BART propensity 
(BART-P) prediction estimator $\widehat{Q}_{ \textrm{BART-P}}$  and the SBART propensity (SBART-P) prediction estimator  $\widehat{Q}_{ \textrm{SBART-P}}$ are outlined as follows. \begin{description}
\item[Step 1] Model $p(I | \boldsymbol{Z}, \boldsymbol{X})$ with probit BART  and estimate $\hat{\pi}$ using posterior mean
\item[Step 2] Obtain the (S)BART-P prediction estimator for finite population mean
    \begin{itemize}
        \item model $p(Y | \boldsymbol{Z}, \boldsymbol{X}, \hat{\pi})$ using (S)BART, $Y = G(\boldsymbol{Z}, \boldsymbol{X}, \hat{\pi}) + \epsilon , \epsilon \sim N(0, \sigma^2)$ 
        \item estimate $\tilde{\theta}_i^{(t)} = G^{(t)} (\boldsymbol{Z}_i, \boldsymbol{X}_i, \hat{\pi}_i)$
        \item $\widehat{Q}_{ \textrm{(S)BART-P}}^{(t)} = \frac{1}{N} \left[ \sum_{i \in U}  \tilde{\theta}_i^{(t)}   + \left (\sum_{i \in s} y_i - \sum_{i \in s} \tilde{\theta}_i^{(t)} \right ) \right]$
        \item $\widehat{Q}_{ \textrm{(S)BART-P}}$: point estimates using (posterior) median of $\widehat{Q}_{ \textrm{(S)BART-P}}^{(t)}$ with credible intervals constructed using quantiles splitting the tails of posterior distribution equally.
    \end{itemize}
\end{description}
BART-P and SBART-P prediction methods are expected to be doubly robust \citep*{long2012doubly}. More specifically, as long as either of the mean model for the outcome or the propensity model is correctly specified, a consistent estimator of the population mean is obtained.

\section{Simulation Studies}
  
\subsection{Simulation Design}\label{sec:design}

Artificial populations with size $N = 3,000$ were simulated. For each unit $i$ in the population, a total number of $p$ binary auxiliary variables and $r$ continuous variables were generated. The $p$ binary variables $\{Z_{il} \}_{l = 1, \ldots p}$ were obtained with $Z_{il} = I(W_{il} < U_l)$, where  $\{ W_{il} \} \overset{\text{i.i.d}}{\sim} \textrm{N}(0, 1)$ and $U_l \overset{\text{i.i.d}}{\sim} \textrm{U}(-.4, .4)$, so that $\text{Pr}(Z_{il} = 1)$ falls in the range $(.34, .66)$, ${l = 1, \ldots, p}$. The $r$ continuous $\{X_{il} \}_{l = 1, \ldots r}$ were generated independently from $\textrm{U}(0, 1)$. Samples of size $n = 600$ were drawn from the populations with inclusion probability $\pi = \textrm{Pr}(I = 1|\boldsymbol{Z}, \boldsymbol{X})$ as a function of the auxiliary variables $\boldsymbol{X}$ and $\boldsymbol{Z}$. We considered the following four simulation scenarios: 
\begin{description}
\item[S1]  {\bf Low-dimensional auxiliary variables ($\boldsymbol{p = 3, r = 1}$) with higher inclusion propensity at the lower tail of $\boldsymbol{X_1}$.} The outcomes $\{Y_{i}\}_{i=1,\ldots,N}$ were generated using an additive model: $Y = 26.81 - Z_1 - 2 Z_2 - 3.5 Z_3 - 25 (X_1 - .75)^2 + \epsilon, \epsilon \sim \text{N}(0, 3^2)$, and the samples were selected with $\pi \propto \text{logit}^{-1}[-13.66 +  .5 Z_1 +  Z_2 + 1.75 Z_3 + 12.5 (X_1 - .75)^2 ]$. Consequently, units with values of $X_1$ falling between 0.5 and 1 were under-sampled. 

\item[S2] {\bf High-dimensional auxiliary variables ($\boldsymbol{p = 30, r = 10}$) with higher inclusion propensity at the lower tail of $\boldsymbol{X_1}$.} Same $Y$ and $\pi$ models as S1, but add noise auxiliary variables $\{Z_l \}_{l = 4, \ldots, 30}$ and $\{ X_l \}_{l = 2, \ldots, 10}$ that are not associated with $Y$ or $\pi$.

\item[S3] {\bf High-dimensional auxiliary variables ($\boldsymbol{p = 30, r = 10}$) with lower inclusion propensity at the lower tail of $\boldsymbol{X_1}$.} Same as S2, but change the signs of the coefficients in the model for $\pi$ to introduce selection bias in the opposite direction: $\pi \propto \text{logit}^{-1}[4.01 -  .5 Z_1 - Z_2 - 1.75 Z_3 - 12.5 (X_1 - .75)^2 ]$. Consequently, units with small values of $X_1$ were under-sampled, especially among those with $X_1 \le 0.25$.

\item[S4] {\bf High-dimensional auxiliary variables ($\boldsymbol{p = 30, r = 10}$) with interaction and different relevant continuous predictors for $\boldsymbol{Y}$ and $\boldsymbol{\pi}$.} 
The outcomes $\{Y_i\}_{i=1,\ldots,N}$ were generated using $Y = 36.81 - Z_1 - 2 Z_2 - 3.5 Z_3 - 10 Z_1 Z_2 - 9 (X_1 - .75)^2 - 16 Z_3 (X_1 - .75)^2 + \epsilon, \epsilon \sim \text{N}(0, 3^2)$, with samples selected using $\pi \propto \text{logit}^{-1}[3.27 -  .5 Z_1 - Z_2 - 1.75 Z_3 - 2 Z_1 Z_2 - 4 (X_3 - .75)^2 - 3 Z_3 (X_3 - .75)^2 - (X_5 - .75)^2]$. Units at the lower tails of $X_3$ and $X_5$ were under-sampled,
but $X_3$ and $X_5$ were not associated with $Y$.
\end{description}

\begin{figure}
    \centering
    \includegraphics[width=\linewidth]{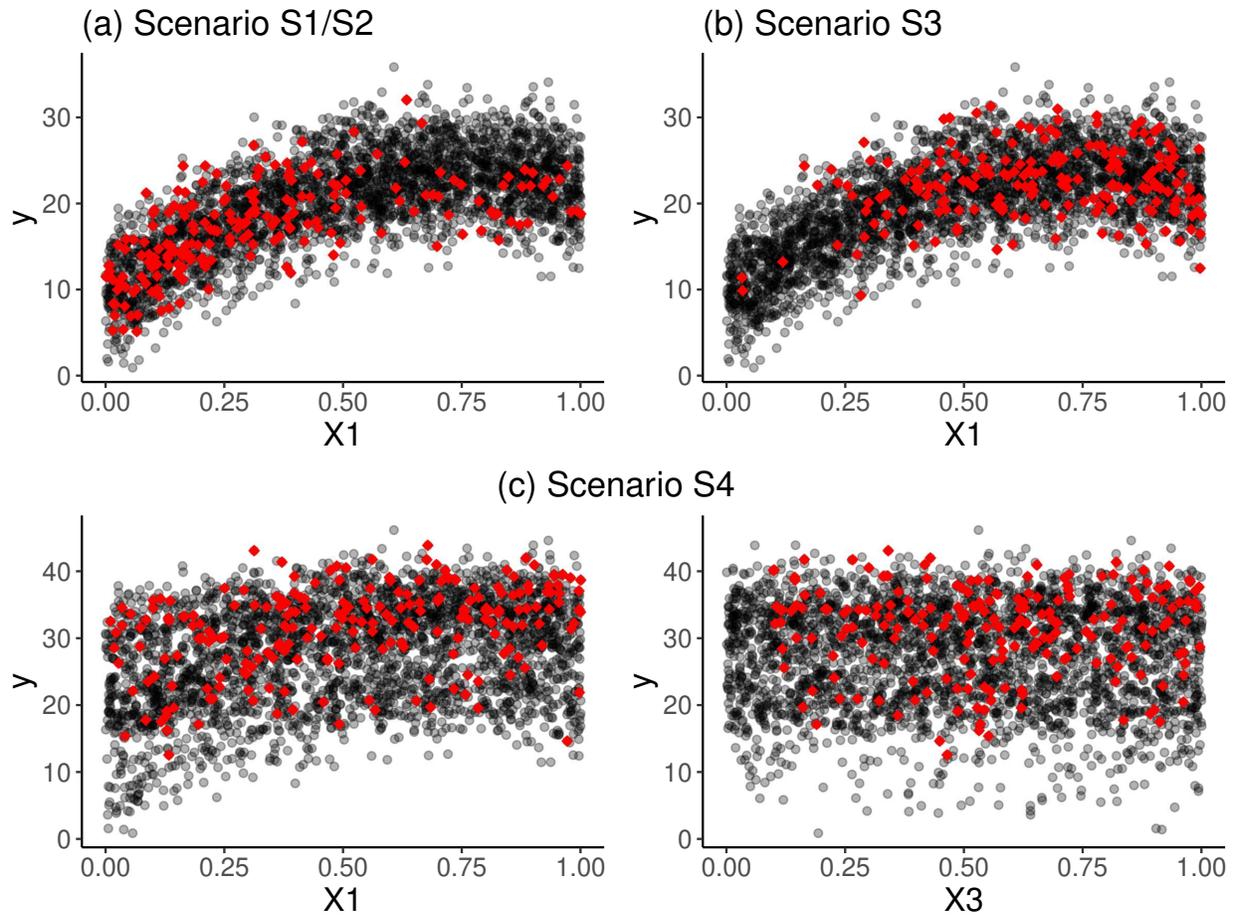}
    \caption{Scatterplots of outcomes $Y$ versus continuous auxiliary variables of units in the population (in gray dots) and a selected sample (in red diamonds) for (a) Scenario S1/S2 (b) Scenario S3 (c) Scenario S4.}
    \label{fig:selection}
\end{figure}

Figure~\ref{fig:selection}(a)-(b) show the scatter plots of $Y$ against $X_1$, the continuous variable that is associated with both $Y$ and $\pi$, of the simulated population overlaid with a selected sample in scenarios S1-S3. Population units with lower values of $X_1$ were more likely to be selected into samples in scenarios S1/S2 but less likely to be selected in scenario S3.

Scenario S4 was designed to assess whether tree-based methods handle interactions well and how they perform when the continuous variables that are associated with undersampling are not associated with outcome. Figure~\ref{fig:selection}(c) visualizes population with a selected sample in scenario S4, using scatter plots of $Y$ against $X_1$, the continuous variable related to $Y$ but not $\pi$, and of $Y$ against $X_3$, the continuous variables related to $\pi$ but not $Y$. The plot on the left shows a positive association between $Y$ and $X_1$ but units with different values of $X_1$ are equally likely to be included in the sample; while the plot on the right shows no association between $Y$ and $X_3$ but units at the lower tail of $X_3$ are less likely to be included in the sample.

For each scenario, 500 replicates of simulation were conducted, with point and interval estimates of finite population mean computed for each. The Bayesian tree-based methods used all available auxiliary variables, as it is unknown which variables are involved in the true data generating process in practice. For scenario S1 with low-dimensional auxiliary variables, the tree-based methods were also compared to the PS and raking estimators using all four available variables with $X_1$ discretized using tertiles in PS and using quintiles in raking. Raw estimates 
were also calculated using sample means.

\subsection{Simulation Results}

\begin{figure}
    \centering
    \includegraphics[width=\linewidth]{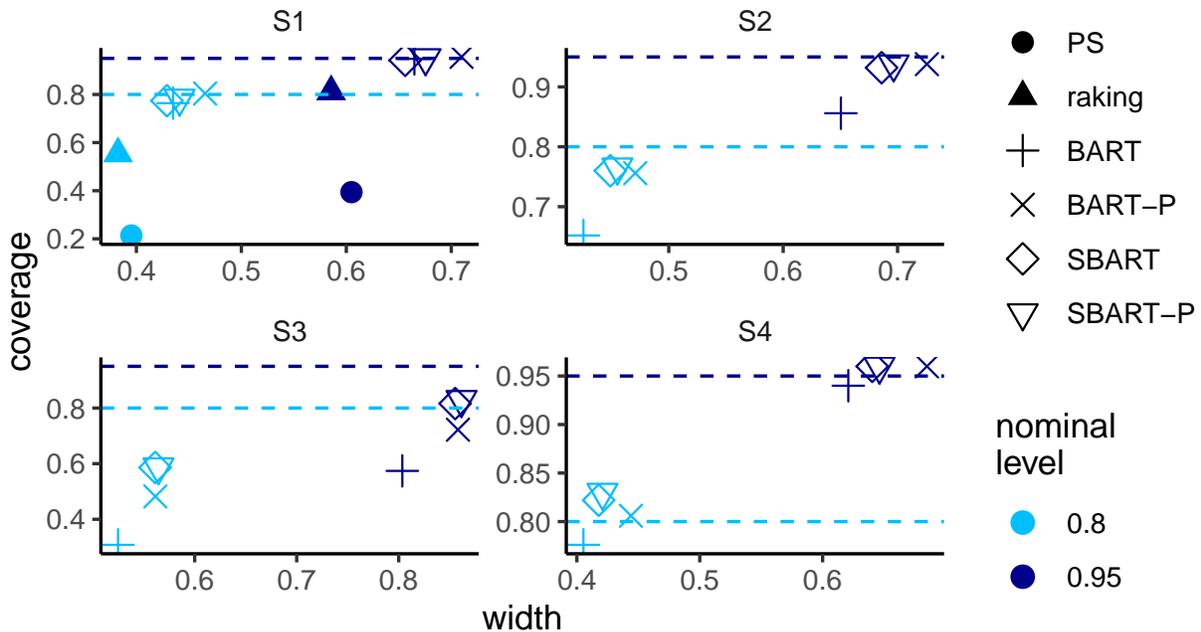}
    \caption{{Simulation results - empirical} coverage rates of $80 \%$ and $95 \%$ probability intervals  (with the horizontal dashed lines denoting the nominal levels) against average probability interval widths, from 500 simulation replicates, for each simulation setting.}
    \label{fig:CI}
\end{figure}

The performance of point estimates are evaluated with empirical bias and empirical root mean squared error (RMSE), summarised in Table~\ref{tab:point}. Scenarios S1-S3 share the same outcome model, the same outcome values $\{Y_i\}_{i=1,\ldots,N}$ and the same ground truth for the finite population mean defined as $Q = \frac{1}{N} \sum_{i=1}^N Y_i$. The empirical coverage rates and average widths of $80\%$ and $95\%$ probability intervals are visualized in Figure~\ref{fig:CI}.   The raw estimates ignoring selection bias are off the chart, leading to confidence intervals with $0\%$ coverage rates, therefore, not shown in Figure~\ref{fig:CI}. 

    \begin{table}[htbp]
        \centering
        \caption{{Simulation results - empirical bias and RMSE of various methods in estimating population means, from 500 simulation replicates, for each simulation setting}}
        \begin{threeparttable} 
\begin{tabular}{lcccccccc} \hline\hline
\multirow{3}{*}{Method} & \multicolumn{2}{c}{S1} &     \multicolumn{2}{c}{S2} & \multicolumn{2}{c}{S3} & \multicolumn{2}{c}{S4}    \\  
\cmidrule(l){2-7} \cmidrule(l){8-9}
        & \multicolumn{6}{c}{$Q =  19.88$} & \multicolumn{2}{c}{$Q =  27.74$} \\
        \cmidrule(l){2-7} \cmidrule(l){8-9}
            & Bias	    &    RMSE   &  Bias	    &    RMSE   &  Bias	    &    RMSE   & Bias	    &    RMSE  \\ \hline
raw	    	& $-2.99$ &	$2.99$ &  $-2.99$& $2.99$	& $2.43$	& $2.43$	& $3.13$	& $3.14$ \\
PS*	    	& $-0.37$ &	$0.43$ &   		& 			&           &           &           &          \\
raking**	& $-0.16$ &	$0.22$ &   		& 			&           &           &           &          \\
BART		& $-0.08$ &	$0.17$ &  $-0.17$ & $0.22$	& $0.37$	& $0.43$	& $0.07$	& $0.17$  \\
BART-P		& $-0.06$ &	$0.18$ &  $-0.12$ & $0.20$	& $0.30$	& $0.38$	& $0.06$	& $0.17$  \\
SBART		& $-0.08$ &	$0.17$ &  $-0.10$ & $0.19$	& $0.24$	& $0.32$	& $0.04$	& $0.16$  \\ 
SBART-P 	& $-0.07$ &	$0.18$ &  $-0.10$ & $0.19$	& $0.24$	& $0.32$	& $0.04$	& $0.16$ \\ \hline
\end{tabular}
\begin{tablenotes}
\footnotesize
\item *PS is based on $Z_1$, $Z_2$, $Z_3$ and $X_1$ discretized using tertiles
\item **Raking is based on $Z_1$, $Z_2$, $Z_3$ and $X_1$ discretized using quintiles.
\item The standard errors of empirical bias from 500 simulation replicates are $< 7.5 \times 10^{-3}$ for all methods  
\end{tablenotes}
\end{threeparttable}
        \label{tab:point}
    \end{table}

In scenario S1, where the weighting methods are feasible, raking is less biased as well as more efficient than post-stratification (PS). This is because raking maintains more information from the continuous variable $X_1$ by discretizing $X_1$ using quintiles as compared to tertiles in PS, and raking implicitly assumes an additive propensity model while PS assumes an interaction model. Both PS and raking generate confidence intervals with coverage rates lower than the nominal levels, with raking yielding shorter intervals but higher coverage rates. BART and SBART both outperform the weighting methods via utilizing the continuous form of $X_1$, generating credible intervals with coverage rates close to the nominal levels. BART and SBART perform similarly as all auxiliary variables are relevant in this low-dimensional setting. Including propensity score in BART and SBART leads to a small bias reduction which is offset by efficiency loss, indicated by slightly higher RMSE and slightly wider credible intervals. 

Scenario S2 differs from scenario S1 by adding irrelevant auxiliary variables. PS and raking are not feasible due to high-dimensionality. Units with $X_1$ falling between 0.5 and 1.0 have lower selection probabilities than those with $X_1$ in between 0 and 0.5 as shown in Figure~\ref{fig:selection}(a). SBART outperforms BART with lower bias, lower RMSE and better credible interval coverage. Including propensity score in BART reduces bias and RMSE and fixes credible interval coverage. However, such improvement is not obvious for SBART. BART-P, SBART and SBART-P all yields valid credible intervals but not BART, with SBART having the shortest intervals.

Scenario S3 differs from scenario S2 in the direction of selection bias. Moreover, because $\pi$ was negatively associated with $(X_1-.75)^2$, the units in the lower tail of $X_1$ (e.g. $X_1 < .25$ in Figure~\ref{fig:selection}(b)) have even smaller inclusion probabilities than the units in the upper tail of $X_1$ in scenario S2. Consequently, there are sparse data in the lower tail of $X_1$. In this setting, neither BART nor SBART performs well with large bias and RMSE, although SBART yields smaller bias and RMSE than BART. The empirical coverage rates for both BART and SBART are lower than the nominal levels due to bias in the estimation. By including propensity score,  BART-P improves credible interval coverage as well as bias and RMSE than BART, but does not fix the undercoverage issue. Again, such improvement is not obvious for SBART. 

In scenario S4, both BART and SBART performs well with small bias and RMSE and close to nominal level coverage rate. SBART yields slightly smaller bias, smaller RMSE, and better coverage rate than BART. Including propensity score slightly reduces bias and improves coverage rate in BART. Although there are sparse data in the lower tails of $X_3$ and $X_5$, these two $X$ variables are not associated with $Y$ and thus such biased selection did not yield poor performance of the tree-based methods like in Scenario S3. 

In all the scenarios considered here, SBART outperforms other competing methods and is, therefore, recommended. However, it should be used with caution, as it still does not perform well when selection bias results in sparse data at the tails of continuous auxiliary variables associated with the outcome.

\subsection{Comparison of BART and SBART Prediction}

\begin{figure}
    \centering
     \includegraphics[width=\linewidth]{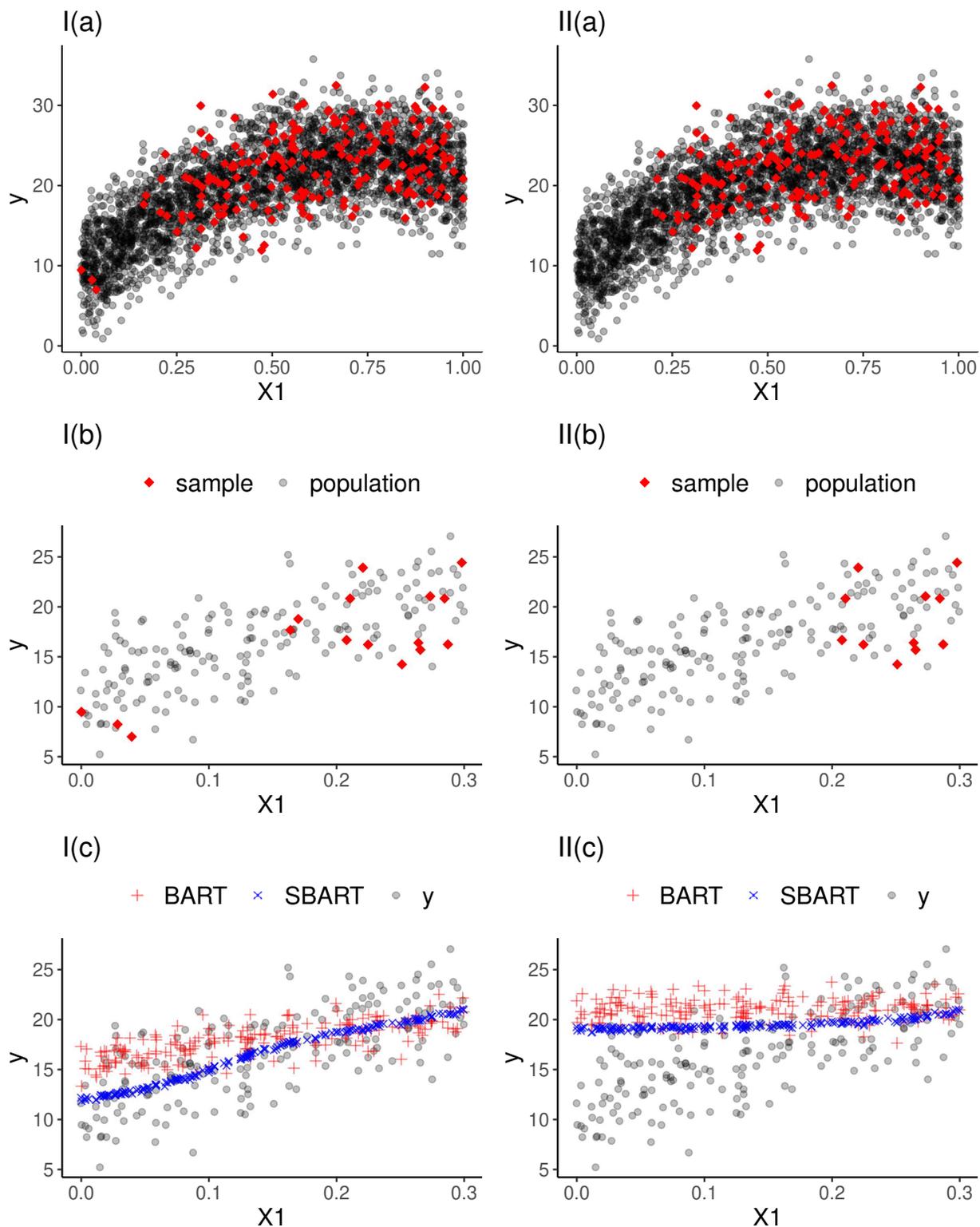}
    \caption{{Two selected samples I and II from the population in Scenario S3: (a) Scatter plots of $Y$ versus $X_1$ with the population in gray dots and a selected sample in red diamonds (b) Scatter plots of $Y$ versus $X_1$, restricted to $Z_2 = Z_3 = 0$ and $X_1 < .3$ (c) Scatter plots of $Y$ versus $X_1$ in the subpopulation, overlapped with posterior means of $G(\boldsymbol{Z},\boldsymbol{X})$ estimated from the BART and SBART models based on the whole sample.}}
    \label{fig:scn_Ic_pred}
\end{figure}

We took a further investigation to compare the performance of BART and SBART in scenario S3, where neither BART nor SBART performs well with BART performing worse than SBART.  We consider two random samples from the population. For sample I, data is sparse at the lower tail of $X_1$, while, for sample II, no data is available at the lower tail, $X_1 < .2$. The top panels I(a) and II(a) of Figure~\ref{fig:scn_Ic_pred} shows the population in gray dots, with sample I and II in red, respectively. For a closer examination of the data at the lower tail of $X_1$, we restrict to a subset with $Z_2 = Z_3 =0$ and focus on the lower tail with $X_1 < 0.3$. The middle panels I(b) and II(b) show the population and corresponding sample data in this restricted subgroup. Finally, the bottom panels I(c) and II(c) plot the population units of $Y$ in this subgroup (gray points) overlaid by the posterior means of the location parameters, $G(\boldsymbol{Z},\boldsymbol{X})$, of each population units estimated using BART and SBART as shown using red pluses and blue crosses, respectively.  Panel I(c) shows that both BART and SBART fit the data well within the region $X_1 > .2$ where sample data are available. However, the SBART fit the data much better than BART when $X_1 < 0.2$ where very sparse data are available. The estimated posterior means of the location parameters based on SBART are also less noisy, due to the sparsity-induced priors that tend to exclude the noise auxiliary variables in model fitting.  Panel II(c) shows that both models fail to produce valid predictions in the region $X_1 < .2$ where there is no sample data available. In the simulation study, about $5\%$ of the 500 simulated samples do not include units with $X_1 < .1$ and one third include fewer than 10 units with $X_1 < .2$.

\section{Applied Examples}\label{sec:app}

We demonstrate the application of the proposed methods using real data from two different studies. The first application example deals with a mental health survey assessing psychiatric disorders among the Ohio Army National Guard (OHARNG) service members. The second application is in a clinic setting where it is of interest to generalize inference on COVID-19 patients when clinical outcomes are only available in a subset of patients.

\subsection{Ohio Army National Guard Survey of Mental Health}

The Ohio Army National Guard (OHARNG) Mental Health Initiative is a population-based observational survey study for estimating the prevalence and identifying correlates of mental illness and health service utilization among the OHARNG service members.  The study population of the baseline survey is defined as all $N = 12570$ soldiers who served in the OHARNG between June 2008 and February 2009. A survey sample with $n = 2562$ service members was selected. In this analysis, we are interested in estimating the mean trauma score among the OHARNG service members using the selected sample, with potential selection bias due to under-coverage of sampling frame and non-response. Auxiliary information is available at individual level for the entire study population, including age (17-24 yr, 25-34 yr, 35+ yr), sex (male, female), race (white, black, other), rank (enlisted, officer), marital status (married, non-married), and years of service (in years).  We apply the proposed trees-based methods to correct the discrepancy between the sample and population utilizing the five categorical and one continuous auxiliary variables. For BART-P and SBART-P, the propensity models were built using probit BART.  

Before modeling, $\log(y + 1)$ transformation was applied to trauma scores to reduce right skewness such that the normality assumption in BART and SBART is reasonable. Distributions of the only continuous variable, years of service, in the sample and population were checked to avoid prediction failure due to sparse data at the tails (see Figure S1 in Supplementary Materials). After fitting the models, we performed model checking using posterior predictive graphics checking \citep[chapter 6]{gelman2014bayesian} based on the following test quantities, including (a) $T_1 (\boldsymbol{y}) = \bar{y}$, (b) $T_2 (\boldsymbol{y}) = \frac{1}{n - 1} \sum_{i = 1}^n (y_i - \bar{y})^2$, and
(c) $T_3(\boldsymbol{y}, G, \sigma) = \frac{1}{n} \sum_{i = 1}^n \left( \frac{y_i - \theta_i}{\sigma} \right)^2$, where $\theta_i = G (\boldsymbol{z}_i, \boldsymbol{x}_i)$. The test quantities catch different aspects of the data, with $T_1(\cdot)$ and $T_2(\cdot)$ measuring the location and variability of the survey outcome while $T_3(\cdot)$ measuring the discrepancy between the survey outcome and fitted distribution. In each MCMC iteration $t$, the realized test quantities $T_i(\boldsymbol{y}, G^{(t)}, \sigma^{(t)})$ under the observed data  and predictive test quantities $T_i(\boldsymbol{\tilde{y}}^{(t)}, G^{(t)}, \sigma^{(t)})$  under the simulated data were computed and compared, with $\boldsymbol{\tilde{y}}^{(t)}$ drawn from the posterior predictive distribution. For each quantity $T_i (\cdot)$,  a Bayesian posterior predictive $p$-value can also be computed, which is defined as the probability that the predictive test quantity is greater than the realized test quantity, evaluated over the posterior distribution. The Bayesian $p$-value measures the discrepancy between the observed data and the posterior predictive distribution in the aspect characterized by $T(\cdot)$. A Bayesian $p$-value close to 0.5 indicates good fit while a Bayesian $p$-value near 0 or 1 indicates that the observed pattern would be unlikely to happen if the model were true and, therefore, lack of fit. Figure S2 in the Supplementary Materials shows the posterior predictive graphics checking and corresponding $p$-values for SBART.  All four Bayesian methods, BART, BART-P, SBART and SBART-P, yielded fitted models with posterior predictive $p$-values close to 0.5, indicating adequate model fit.

\begin{figure}
    \centering
    \includegraphics[width=\linewidth]{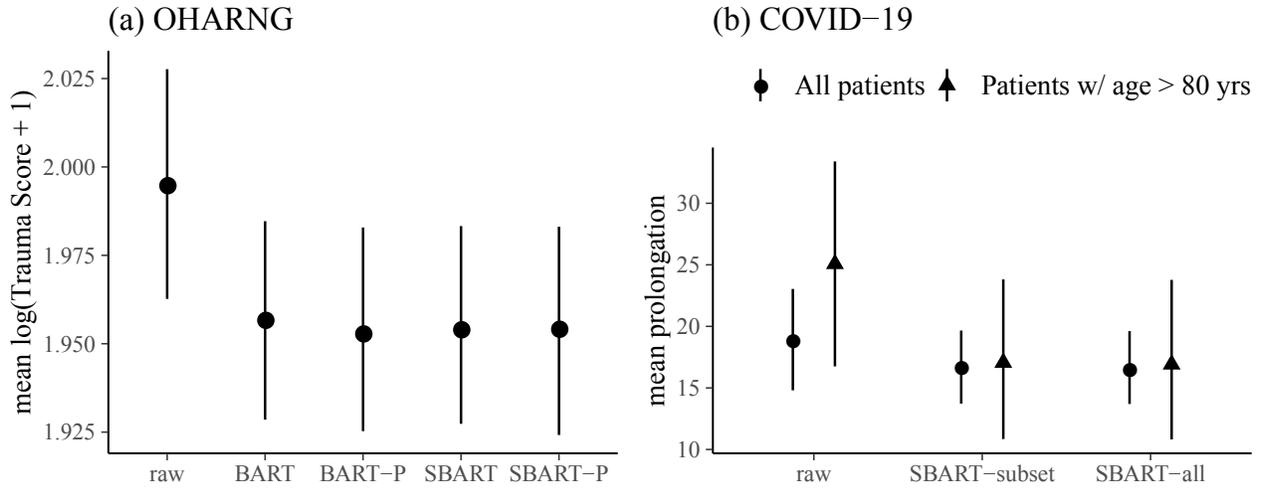}
    \caption{(a) Point estimates and $95 \%$ probability intervals of mean $\log$(trauma score + 1) among soldiers who served in the OHARNG between June 2008 and February 2009 (b) Point estimates and $95 \%$ probability intervals of mean prolongation among all patients and patients with age $\geq 80$ years old, comparing raw sample means, SBART with baseline QTc and treatment (SBART-subset), and SBART with all auxiliary variables (SBART-all).}
    \label{fig:app}
\end{figure}

We compare the results of proposed Bayesian methods with the raw estimates in estimating the mean trauma score on the log scale, with point estimates and $95 \%$ probability intervals visualized in Figure~\ref{fig:app}(a). The Bayesian methods yields lower estimates for mean trauma score compared to raw estimates without adjustment. BART and SBART yields similar results, as this is a low-dimensional setting with one continuous auxiliary variable and the benefit of soft decision trees is not so obvious. Including propensity scores do not lead to much change in the estimates. We recommend reporting the estimates using SBART in this analysis.

\subsection{New York City COVID-19 Study}

The COVID-19 is a global pandemic caused by severe acute respiratory syndrome coronavirus 2 (SARS-Cov-2). The first positive case was confirmed in New York City on March 1, 2020. The city had more cases than any country other than the United States by May 2020. The urgent need for therapeutic agents has resulted in repurposing and redeployment of experimental agents. Hydroxychloroquine, combined with azithromycin, was administered to patients with COVID-19 without robust evidence supporting its use, with the U.S. Food and Drug Administration (FDA) issuing an emergency use authorization (EUA) to allow doctors to begin treating patients with hydroxychloroquine in hospitalized settings outside clinical trials on March 28, 2020. Such EUA was later revoked as of June 15, 2020, with a randomized clinical trial in hospitalized patients showing no benefits and reports of serious heart rhythm problems along with other safety issues. Both hydroxychloroquine and azithromycin are characterized as definite QTc prolongers that increase risk of sudden cardiac deaths. Between March 1st, 2020 through May 1st, 2020, there were 470 patients admitted to Columbia University Irving Medical Center, treated with hydroxychloroquine (H+) or hydroxychloroquine combined with azithromycin (A+H+) \citep{rubin2021}. All patients have baseline ECG measurements of QTc while, but only 244 of them have ECG QTc measurements on Day 2 of medication. We are interested in estimating the mean QTc prolongation, defined as difference in QTc measures between day 2 and day 0, of all the 470 COVID-19 patients who received H+ or A+H+ treatments. However, the QTc prolongation was only measured among the 244 patients who had QTc measurements at day 2. To improve the estimation, we also collected the data of these 470 patients on their demographic characteristics and relevant biomarkers from electronic medical records.

Exploratory analysis indicates a strong negative  association between prolongation and baseline QTc measurement and that patients with higher baseline QTc are less likely to have ECG QTc measurement on Day 2, demonstrated in Figure S3 in the Supplementary Materials.  Other auxiliary variables include treatment (H+, A+H+), demographic characteristics, including gender, age (in years), race (white, black, other), BMI (log scale), along with 7 biomarkers. We used the recommended method SBART for two estimands of interest: (i) the mean QTc prolongation among all 470 patients, and (ii) the mean QTc prolongation among the 87 (out of 470) patients who were over 80 years old. We compared two SBART models, with the first model only including baseline QTc and treatment (SBART-subset) and the second model including all covariates (SBART-all).

As is visualized in Figure~\ref{fig:app}(b), SBART yields lower estimates of mean QTc prolongation compared to the raw estimates ignoring selection bias, for both estimands of interest. For estimand (i), including baseline QTc and treatment in SBART leads to obvious drop in the mean prolongation estimates {from a raw estimate of $18.9~(95\% \text{~CI}: 14.8, 23.0)$ milliseconds to $16.7~(95\% \text{~CI}: 13.7, 19.7)$ milliseconds}. Additionally adding other auxiliary variables does not lead to further obvious change in the estimates.

For estimand (ii), estimates can be readily obtained by restricting the calculation to predicted and observed prolongation for patients over 80 years old, without additional computation for model fitting. Similar to the estimate among all patients, SBART yields smaller estimates and shorter probability intervals than the raw estimator. Although the mean prolongation estimate is higher among this subgroup of patients than all patients using the raw estimator, the estimates are similar using SBART.

\section{Discussion}
\label{sec:discuss}

We consider generalization of inference on a descriptive estimand from a non-random sample to a target population in data-rich settings where high dimensional auxiliary information is available in both the sample and population, with survey inference being a special case. Existing methods such as post-stratification, raking and MRP are challenging or infeasible to be performed due to high-dimensionality and the need to discretize continuous auxiliary variables before applying such methods leads to loss of information. To address such issues, we propose a regularized prediction approach by modeling the conditional distribution of the outcomes given the high-dimensional auxiliary variables using Bayesian machine learning techniques. In this paper, we specifically consider BART and soft BART which handles both discrete and continuous auxiliary variables as well as potential interactions. Besides the auxiliary variables, we also {consider} modified methods that estimates the propensity score for a unit to be included in the sample and also include the estimated propensity score as a covariate in the BART and soft BART model.

Artificial data simulation studies demonstrate that the Bayesian additive-tree{s}-based methods outperform post-stratification (PS) and raking in low-dimensional settings where PS and raking are feasible, as the regularized additive trees better utilize information in the continuous auxiliary variables and avoid model overspecification.  The Bayesian additive-trees-based methods also yield valid inference in high-dimensional settings when PS and raking are not feasible, as long as selection bias does not result in sparse data points at the tails of relevant continuous auxiliary variables. In high-dimensional setting with sparse signals, SBART, with soft decision trees and sparsity-inducing priors, is less biased and more efficient than BART. In challenging settings where the additive-trees-based methods underperform, including propensity score in BART could reduce bias and improve credible interval coverage while such benefit is not obvious for SBART. Therefore, the soft BART prediction method is recommended for generalization of inference with high-dimensional auxiliary variables. The soft BART better utilizes information in the continuous auxiliary variables and more effectively regularize the effect of irrelevant noise auxiliary variable. 
As is demonstrated in the OHARNG mental health study and the COVID-19 study, the proposed methods could be applied in both survey and more general settings, with estimands being overall population as well as subpopulation quantities.

The Bayesian additive-trees-based methods need to be used with caution. More specifically, both BART and SBART prediction fail when selection bias results in very sparse data point at the tails of the continuous covariates. Such prediction failure cannot be fixed via robust methods involving propensity scores. Therefore, for important continuous variables associated with the outcomes, the range and distribution in the sample and population need to be checked before using the methods. In some cases, transformation on such auxiliary variables could be applied to reduce sparsity at the tails.

Although BART and SBART are considered in this paper, the regularized prediction approach is general and any Bayesian machine learning techniques that achieve valid predictions could be applied.




\section*{Acknowledgements}

This work was partially supported by NIH grants R01AG067149 and R21ES029668 and ONR grant N00014-17-1-2141. The authors thank Dr.\ Sandro Galea for sharing the data on the OHARNG service members, Drs.\ Elaine Wan and Marc Waase for sharing the data on COVID-19 patients admitted at Columbia University Irving Medical Center. \vspace*{-8pt}




\bibliographystyle{agsm}
\bibliography{RP-BART-ref}

@article{wang2015forecasting,
  title={Forecasting elections with non-representative polls},
  author={Wang, Wei and Rothschild, David and Goel, Sharad and Gelman, Andrew},
  journal={International Journal of Forecasting},
  volume={31},
  number={3},
  pages={980--991},
  year={2015},
  publisher={Elsevier}
}

@article{gelman2007struggles,
  title={Struggles with survey weighting and regression modeling},
  author={Gelman, Andrew},
  journal={Statistical Science},
  volume={22},
  number={2},
  pages={153--164},
  year={2007},
  publisher={JSTOR}
}

@article{gelman1997poststratification,
  title={Poststratification into many categories using hierarchical logistic regression},
  author={Gelman, Andrew and Little, Thomas C},
  year={1997},
  journal={Survey Methodology},
  volume={23},
  number={2},
  pages={127--135},
  publisher={Citeseer}
}

@article{little2004robust,
  title={Robust likelihood-based analysis of multivariate data with missing values},
  author={Little, Roderick and An, Hyonggin},
  journal={Statistica Sinica},
  volume={14},
  number={3},
  pages={949--968},
  year={2004},
  publisher={JSTOR}
}

@article{chipman2010bart,
  title={BART: Bayesian additive regression trees},
  author={Chipman, Hugh A and George, Edward I and McCulloch, Robert E},
  journal={The Annals of Applied Statistics},
  volume={4},
  number={1},
  pages={266--298},
  year={2010},
  publisher={Institute of Mathematical Statistics}
}

@article{linero2018bayesian,
  title={Bayesian regression tree ensembles that adapt to smoothness and sparsity},
  author={Linero, Antonio R and Yang, Yun},
  journal={Journal of the Royal Statistical Society: Series B (Statistical Methodology)},
  volume={80},
  number={5},
  pages={1087--1110},
  year={2018},
  publisher={Wiley Online Library}
}

@article{tan2019robust,
  title={“Robust-Squared” Imputation Models using Bart},
  author={Tan, Yaoyuan V and Flannagan, Carol AC and Elliott, Michael R},
  journal={Journal of Survey Statistics and Methodology},
  volume={7},
  number={4},
  pages={465--497},
  year={2019},
  publisher={Oxford University Press}
}

@book{gelman2014bayesian,
  title={Bayesian data analysis},
  author={Gelman, Andrew and Carlin, John B and Stern, Hal S and Dunson, David B and Vehtari, Aki and Rubin, Donald B},
  volume={},
  year={2014},
  publisher={CRC press}
}

@article{rothwell2005external,
  title={External validity of randomised controlled trials:“to whom do the results of this trial apply?”},
  author={Rothwell, Peter M},
  journal={The Lancet},
  volume={365},
  number={9453},
  pages={82--93},
  year={2005},
  publisher={Elsevier}
}

@article{hill2011bayesian,
  title={Bayesian nonparametric modeling for causal inference},
  author={Hill, Jennifer L},
  journal={Journal of Computational and Graphical Statistics},
  volume={20},
  number={1},
  pages={217--240},
  year={2011},
  publisher={Taylor \& Francis}
}

@article{valliant1993poststratification,
  title={Poststratification and conditional variance estimation},
  author={Valliant, Richard},
  journal={Journal of the American Statistical Association},
  volume={88},
  number={421},
  pages={89--96},
  year={1993},
  publisher={Taylor \& Francis Group}
}

@article{deming1940least,
  title={On a least squares adjustment of a sampled frequency table when the expected marginal totals are known},
  author={Deming, W Edwards and Stephan, Frederick F},
  journal={The Annals of Mathematical Statistics},
  volume={11},
  number={4},
  pages={427--444},
  year={1940},
  publisher={JSTOR}
}

@article{hahn2020bayesian,
  title={Bayesian regression tree models for causal inference: Regularization, confounding, and heterogeneous effects (with discussion)},
  author={Hahn, P Richard and Murray, Jared S and Carvalho, Carlos M},
  journal={Bayesian Analysis},
  volume={15},
  number={3},
  pages={965--1056},
  year={2020},
  publisher={International Society for Bayesian Analysis}
}

@article{long2012doubly,
  title={Doubly robust nonparametric multiple imputation for ignorable missing data},
  author={Long, Qi and Hsu, Chiu-Hsieh and Li, Yisheng},
  journal={Statistica Sinica},
  volume={22},
  pages={149},
  year={2012},
  publisher={NIH Public Access}
}

@article{rubin2021,
  title={COVID-19 infection is associated with QTc prolongation},
  author={Geoffrey A. Rubin and Amar D. Desai and Zilan Chai and Aijin Wang and  Qixuan Chen and Amy S. Wang and Cameron Kemal and Haajra Baksh and Angelo Biviano and Jose M. Dizon and Hirad Yarmohammadi and Frederick Ehlert and Deepak Saluja and David A. Rubin and John P. Morrow and Uma Mahesh R. Avula and Jeremy P. Berman and Alexander Kushnir and Mark P. Abrams and Jessica A. Hennessey and  Pierre Elias and Timothy J. Poterucha and Nir Uriel and Christine J. Kubin and  Elijah LaSota and Jason Zucker and Magdalena E. Sobieszczyk and Allan Schwartz and Hasan Garan and Marc P. Waase and  and Elaine Y. Wan},
  journal={JAMA Network Open, accepted},
  volume={},
  pages={},
  year={2021+},
  publisher={}
}

\end{document}